\documentstyle[prl,aps,preprint,floats]{revtex}

\begin{document}

\title{Plane waves in metric-affine gravity}

\author{Yuri N.~Obukhov\footnote{On leave from: 
Dept. of Theoret. Physics, Moscow State University, 117234 Moscow, Russia}}
\address{Institute for Theoretical Physics, University of Cologne,
50923 K\"oln, Germany}

\maketitle

\begin{abstract}
We describe plane-fronted waves in the Yang-Mills type quadratic metric-affine
theory of gravity. The torsion and the nonmetricity are both nontrivial, and
they do not belong to the triplet ansatz. 
\end{abstract}
\bigskip


\section{Introduction}

Although Einstein's general relativity theory is satisfactorily supported 
by experimental tests on a macroscopic level, the gravitational interaction
on a microscopic scale is not well understood. The gravitational gauge 
models provide an alternative description of gravitational physics in 
the microworld \cite{PR}. A variety of models arise within the framework of 
the gauge approach to gravity (Poincar\'e, teleparallel, metric-affine, 
supergravity, to mention but a few), and their corresponding {\it kinematic} 
schemes are well established at present. However, the {\it dynamic} aspects 
of the gauge gravity models have been rather poorly studied up to now. This
includes the choice of the basic Lagrangian of the theory, as well as the 
detailed analysis of possible physical effects. The derivation of a new
exact solutions for these models may bring new insight to the understanding
of gravitational physics on small scales. 

The plane-fronted gravitational waves represent an important class of exact 
solutions which generalize the basic properties of electromagnetic waves
in flat spacetime to the case of curved spacetime geometry. The relevant
investigation of the gravitational waves in general relativity has a
long and rich history, see, e.g., \cite{peres,pen1,pen2,griff,vdz,exact}.
The discussion of the possible generalizations of such solutions revealed
the exact wave solutions in Poincar\'e gauge gravity 
\cite{adam,chen,sippel,vadim,singh,babu}, in teleparallel gravity 
\cite{tele}, in generalized Einstein theories \cite{gurses,lovelock}, 
in supergravity \cite{sg1,sg2,sg3,sg4,sg5}, as well as, more recently, 
in superstring theories \cite{gimon,ark1,ark2,str1,str2,str3,str4}.
Some attention has also been paid to the higher-dimensional generalizations
of the gravitational wave solutions \cite{coley1,coley2,hervik,ndim}. 
It was demonstrated \cite{dirk1,dirk2,king,vas1,vas2,vas3} that gravitational 
wave solutions are also admitted in the metric-affine theory of gravity (MAG) 
with the propagating torsion and nonmetricity fields. The latter results are, 
however, restricted either to the case of torsion waves only, or to the 
triplet class of solutions with a specific ansatz for torsion and
nonmetricity \cite{eff,tri} and for a special form of the Lagrangian. 

The aim of this paper is to describe the plane gravitational waves for the 
general Yang-Mills type quadratic MAG Lagrangian with nontrivial torsion and 
nonmetricity configurations that do not belong to the triplet ansatz. 
The motivation is twofold. On the one hand, the systematic study of the
space of solutions represents a significant aspect of the development of
any field-theoretic model. On the other hand, the wave phenomena as such 
are of fundamental importance, and the construction and comparison of the
wave solutions in different models may clarify the physical contents of and 
the relations between the microscopic and macroscopic gravitational theories
(in particular, general relativity, Poincar\'e gauge gravity and MAG).  

The metric-affine spacetime is described by the metric
$g_{\alpha\beta}$, the coframe 1-forms $\vartheta^{\alpha}$, and the linear
connection 1-forms $\Gamma_{\beta}{}^{\alpha}$. These are interpreted as
generalized gauge potentials, while the corresponding field strengths
are the nonmetricity 1-form $Q_{\alpha\beta}=-Dg_{\alpha\beta}$ and the
2-forms of torsion $T^{\alpha}=D\vartheta^{\alpha}$ and curvature 
$R_{\beta}{}^{\alpha}=d\Gamma_{\beta}{}^{\alpha} + \Gamma_{\gamma}{}^{\alpha}
\wedge\Gamma_{\beta}{}^{\gamma}$. The metric-affine geometry reduces to a 
purely Riemannian one as soon as torsion and nonmetricity both vanish. The 
teleparallel geometry arises when the curvature is trivial, $R_{\beta}
{}^{\alpha} = 0$, whereas a vanishing nonmetricity $Q_{\alpha\beta}=0$ 
yields the Riemann-Cartan geometry of spacetime. It is well known that for 
every metric $g_{\alpha\beta}$ there exists a unique torsion-free and 
metric-compatible connection represented by the Christoffel symbols. We 
will denote this Riemannian connection by $\widetilde{\Gamma}_{\beta}
{}^{\alpha}$, and hereafter the tilde will denote purely Riemannian 
geometrical objects and covariant differentials constructed from them. 
Our general notations and conventions for the basic geometric objects, 
the holonomic and anholonomic indices, the choice of the metric signature 
are that of \cite{PR}. 

The plan of the paper is as follows. In the next Sec.~\ref{emwave}, we 
recall the definition of the ordinary electromagnetic wave. This is used then
in Sec.~\ref{ansatz} for the description of the corresponding ansatz for
a gravitational plane wave in MAG. The properties of the resulting 
curvature, torsion and nonmetricity are discussed in Sec.~\ref{NTC}. 
Finally, in Sec.~\ref{eqs} we demonstrate that the proposed ansatz provides
the exact solution for the general quadratic MAG model. The conclusions are
outlined in Sec~\ref{conclusion}.

\section{Electromagnetic plane waves}\label{emwave}

An electromagnetic plane wave is described by a 1-form $u$ which 
satisfies 
\begin{eqnarray}
d\,{}^\ast du &=& 0,\label{ddu}\\
k\wedge{}^\ast du = 0,\qquad k\wedge du &=& 0.\label{kdu}
\end{eqnarray}
The propagation 1-form $k$ is null (i.e., $k\wedge{}^\ast k =0$) and 
geodetic, $k\wedge{}^\ast dk =0$, and it corresponds to a congruence with 
zero shear, expansion and rotation. This is typical for the plane wave, and
the equations (\ref{kdu}) represent the so-called radiation conditions
imposed on the electromagnetic field. With the electromagnetic potential
$A = u$, the electromagnetic field strength $F = du$ satisfies the vacuum 
Maxwell equation (\ref{ddu}). 

We do not specify the spacetime metric (and hence the Hodge operator ${}^\ast$)
as the flat Minkowski one. It will be convenient not to fix the spacetime 
geometry at this stage.
 
Such an electromagnetic wave construction underlies the derivation of the
corresponding gravitational wave solutions as described in
\cite{pleb,orr,pod1,pod2,pod3}, for example. In our study, we will use
the similar constructions by extending the Riemannian results to their
non-Riemannian counterparts.

\section{Wave ansatz for metric, coframe and affine connection}\label{ansatz}

Let us denote the local spacetime coordinates as $x^i = \{\sigma, \rho, z^2, 
z^3\}$. The upper case Latin indices, $A,B,\dots = 0, 1$, will label the first
2 spacetime dimensions which are relevant to a $pp$-wave. In particular, 
$x^A = \{\sigma, \rho\}$ are the wave coordinates with the wave fronts 
described by the surfaces of constant $\sigma$, and $\rho$ is an affine 
parameter along the wave vector of the null geodesic. The lower case Latin 
indices, $a,b,\dots = 2,3$, refer to the remaining spatial coordinates:
$x^a = \{z^2, z^3\}$. The Greek indices, $\alpha, \beta, \dots = 0,\dots, 3$, 
label the local anholonomic (co)frame components. We denote separate frame 
components by a circumflex over the corresponding index in order to 
distinguish them from coordinate components.

The can now formulate the wave ansatz for the MAG gravitational potentials 
$g_{\alpha\beta}$, $\vartheta^{\alpha}$, and $\Gamma_{\beta}{}^{\alpha}$ as 
follows. We choose the half-null metric  
\begin{equation}
g_{\alpha\beta} = \left(\begin{array}{cc}g_{AB}&0\\ 0&g_{ab}\end{array}
\right),\qquad g_{AB} = \left(\begin{array}{cc}0&1\\ 1&0\end{array}\right),
\quad g_{ab} = \delta_{ab}.\label{met}
\end{equation}
The components of the coframe 1-form are given by
\begin{equation}\label{cof}
\vartheta^{\widehat{0}} = -\,d\sigma,\qquad \vartheta^{\widehat{1}} = {\frac 
12}\,H(\sigma, z^a)\,d\sigma + d\rho,\qquad \vartheta^a = dz^a,\quad a=2,3.
\end{equation}
Finally, the ansatz for the affine connection reads:
\begin{equation}\label{conn}
\Gamma^{\alpha\beta} = k^{[\alpha}\varphi^{\beta]}\,k + k^\alpha k^\beta\,u.
\end{equation}
The corresponding dual frame basis (such that $e_\alpha\rfloor\vartheta^\beta
= \delta_\alpha^\beta$) reads:
\begin{equation}
e_{\widehat{0}} = -\,\partial_\sigma + H\,\partial_\rho,\qquad 
e_{\widehat{1}} = \partial_\rho,\qquad e_a = \partial_a.\label{frame}
\end{equation}

We now make a crucial assumption that the 1-forms $u$ and $k$ above fulfill 
the radiation conditions (\ref{ddu}), (\ref{kdu}). Moreover, using the  
local coordinate system adapted for a plane wave and the half-null nature of 
the coframe (\ref{cof}), we can put $k = \vartheta^{\widehat{0}}$ without 
loss of generality. As usual, we introduce the components as $k_\alpha 
= e_\alpha\rfloor k$. Finally, the components of $\varphi_\alpha$ are 
determined by the function $H$ as follows:
\begin{equation}
\varphi_{\widehat{0}} = 0,\qquad \varphi_{\widehat{1}} = 0,\qquad
\varphi_a = \partial_a H. 
\end{equation}
Although this choice looks to be rather ad hoc, it is actually well 
motivated by the corresponding Riemannian solution, cf. \cite{ndim}. 
Indeed, the ansatz (\ref{met}) and (\ref{cof}) for the metric and the 
coframe is exactly the same as in the purely Riemannian case, whereas 
the connection (\ref{conn}) minimally extends the Christoffel connection
(given in Appendix of \cite{ndim}) via the term proportional to the 
non-Riemannian parameter $u$. In the next section we demonstrate that 
both the torsion and the nonmetricity are determined by this 1-form. 

It is worthwhile to note that the wave 1-form is closed, $dk = 0$, whereas 
the wave covector is covariantly constant,
\begin{equation}
D k_\alpha = -\,\Gamma_\alpha{}^\beta\,k_\beta = 0,\qquad
D k^\alpha = \Gamma_\beta{}^\alpha\,k^\beta = 0.\label{Dk}
\end{equation}
Here we used the fact that this covector is null, $k^\alpha k_\alpha =0$
and orthogonal to $\varphi_\alpha$, i.e. $k^\alpha\varphi_\alpha =0$.

\section{Nonmetricity, torsion and curvature}\label{NTC}

Given the above ansatz for the MAG potentials -- metric (\ref{met}), coframe
(\ref{cof}) and linear connection (\ref{conn}) -- it is straightforward to 
find the corresponding gauge field strengths. The nonmetricity, torsion and 
curvature read, explicitly:
\begin{eqnarray}
Q_{\alpha\beta} &=& 2k_\alpha k_\beta\,u,\label{nonm}\\
T^\alpha &=& k^\alpha\,u\wedge k,\label{tor}\\
R^{\alpha\beta} &=& 2\gamma^{[\alpha}k^{\beta]}\wedge k 
+ k^\alpha k^\beta\,du.\label{curv}
\end{eqnarray} 
Here the covector-valued 1-form is defined by $\gamma_\alpha = - {\frac 12}
\,\underline{d}\varphi_\alpha$, where the differential $\underline{d}$ is
taken with respect to the $z^a$ coordinates only. This 1-form has the 
obvious properties: $k^\alpha\gamma_\alpha = 0$, $\vartheta^\alpha\wedge
\gamma_\alpha = 0$ and $e_A\rfloor\gamma_\alpha = 0$, $A = 0,1$. 

When $u = 0$, we find zero torsion and nonmetricity. In this sense, we may
consider the nontrivial $u$ to represent the true post-Riemannian geometric
structures which we are primarily interested in.

Let us compute the irreducible parts of the curvature 2-form. It is well 
known \cite{PR} that the curvature for the general linear connection can
be decomposed into 11 irreducible parts. Following \cite{PR}, we first 
decompose the curvature 2-form into the skew-symmetric and symmetric forms,
\begin{equation}
W^{\alpha\beta} = R^{[\alpha\beta]} = 2\gamma^{[\alpha}k^{\beta]}\wedge k,
\qquad Z^{\alpha\beta} = R^{(\alpha\beta)} = k^\alpha k^\beta\,du.\label{WZ}
\end{equation}
Then we have to calculate the interior products with the frame and exterior 
products with the coframe. In tensor language, this corresponds 
to computing the various contractions of the curvature tensor. 

All the contractions of the symmetric curvature are trivial. Namely, $Z =
Z_\alpha{}^\alpha = 0$ in view of the nullity of the wave vector ($k_\alpha
k^\alpha = 0$), whereas
\begin{equation}
e_\alpha\rfloor Z^{\alpha\beta} = -\,k^\beta\,{}^\ast(k\wedge{}^\ast du) =0,
\qquad \vartheta_\alpha\wedge Z^{\alpha\beta} = k^\beta\,k\wedge du =0,
\end{equation}
due to the fact that $k_\alpha\vartheta^\alpha = k$ and using the radiation 
conditions (\ref{kdu}). As a result, all the irreducible parts of the 
symmetric curvature form are zero except for the first piece:
\begin{equation}
Z_{\alpha\beta} = {}^{(1)}\!Z_{\alpha\beta}.\label{Z1}
\end{equation}
In tensor language this means that all the contractions of the symmetric
part of the curvature tensor are trivial, i.e., this tensor is totally
trace-free and dual trace-free. The symmetric part of the curvature has
no Riemannian counterpart, this is a totally post-Riemannian object. 

For the skew-symmetric curvature we find straightforwardly:
\begin{equation}
e_\alpha\rfloor W^{\alpha\beta} = (e_\alpha\rfloor\gamma^\alpha)\,k^\beta\,k,
\qquad \vartheta_\alpha\wedge W^{\alpha\beta} = 0.
\end{equation}
Accordingly, if we demand that the zero-form $e_\alpha\rfloor\gamma^\alpha$ 
vanishes, then all the contractions of the skew-symmetric curvature are also 
trivial. This condition imposes the partial differential equation on the 
unknown function $H$:
\begin{equation}\label{ddH}
e_\alpha\rfloor\gamma^\alpha = -{\frac 12}\,\partial_a\partial^a H = 0. 
\end{equation}
Provided that $H(\sigma, z^a)$ is a solution of the Laplace equation 
(\ref{ddH}), we ultimately find that all the irreducible parts of the 
skew-symmetric curvature form are zero except for the first piece
\begin{equation}
W_{\alpha\beta} = {}^{(1)}\!W_{\alpha\beta}.\label{W1}
\end{equation}
This is again the pure tensor part which is totally trace-free and 
dual trace-free, in complete analogy with the symmetric curvature. The
2-form ${}^{(1)}\!W_{\alpha\beta}$ is a direct non-Riemannian generalization 
of the Weyl tensor.

\section{Field equations: quadratic MAG model}\label{eqs}

Let us consider the general Yang-Mills type (curvature quadratic) Lagrangian 
for the MAG model which was studied recently in the literature 
(see, for example, \cite{dirk1,dirk2,king,vas1,vas2,vas3}):
\begin{eqnarray}
V_{\rm MAG}&=& -\,{\frac{1}{2}}\,R^{\alpha\beta}
\wedge{}^*\!\left(\sum_{I=1}^{6}w_{I}\,^{(I)}W_{\alpha\beta}
+ w_7\,\vartheta_\alpha\wedge(e_\gamma\rfloor
{}^{(5)}W^\gamma{}_{\beta} ) \nonumber\right.\\
&& +\,\left.\sum_{I=1}^{5}{z}_{I}\,^{(I)}Z_{\alpha\beta} + z_6
\,\vartheta_\gamma\wedge (e_\alpha\rfloor ^{(2)}Z^\gamma{}_{\beta})
+\sum_{I=7}^{9}z_I\,\vartheta_\alpha\wedge(e_\gamma\rfloor
  ^{(I-4)}Z^\gamma{}_{\beta} )\right)\label{QMA}\,. 
\end{eqnarray}
The 16 dimensionless coupling constants $w_1, \ldots w_7$, $z_1, \ldots z_9$
describe the contributions of all possible quadratic invariants which can be
constructed from the components of the curvature in a general MAG theory
\cite{remark}.  

The vacuum gravitational field equations of the MAG theory read \cite{PR}:
\begin{eqnarray} 
  DH_{\alpha}- E_{\alpha}&=& 0,\label{first}\\ 
  DH^{\alpha}{}_{\beta}-E^{\alpha}{}_{\beta}&=& 0.\label{second}
\end{eqnarray}
The gravitational gauge field momenta are introduced by partial 
differentiation,
\begin{equation}  
H_\alpha = - \frac{\partial V}{\partial T^\alpha}\,,\quad 
H^\alpha{}_\beta= -\frac{ \partial V}{\partial R_\alpha{}^\beta}\,,\quad
M^{\alpha\beta} = - 2\frac{\partial V}{\partial Q_{\alpha\beta}}\,,\label{3}
\end{equation}
whereas the canonical gauge field currents of the gravitational 
energy--momentum and of the hypermomentum, respectively, are defined as
the following expressions, linear in the Lagrangian and in the gauge field 
momenta:
\begin{eqnarray}
  E_{\alpha} & := & \frac{\partial V}{\partial\vartheta^\alpha}
  =e_{\alpha}\rfloor V + (e_{\alpha}\rfloor T^{\beta}) \wedge
  H_{\beta} + (e_{\alpha}\rfloor R_{\beta}{}^{\gamma})\wedge
  H^{\beta}{}_{\gamma} + {1\over 2}(e_{\alpha}\rfloor Q_{\beta\gamma})
  M^{\beta\gamma}\,,\\ E^{\alpha}{}_{\beta} & := &\frac{\partial
    V}{\partial\Gamma_\alpha{}^\beta}= - \vartheta^{\alpha}\wedge
  H_{\beta} - g_{\beta\gamma}M^{\alpha\gamma}\,.
\end{eqnarray}

For the purely curvature quadratic Lagrangian (\ref{QMA}), we obviously have 
$H_\alpha =0$ and $M^{\alpha\beta} = 0$. Hence $E^{\alpha}{}_{\beta} = 0$, 
and as a result the field equations (\ref{first}) and (\ref{second}) reduce to
\begin{eqnarray}
E_{\alpha} = e_{\alpha}\rfloor V + (e_{\alpha}\rfloor 
R_{\beta}{}^{\gamma})\wedge H^{\beta}{}_{\gamma} &=& 0,\label{first2}\\
D H^\alpha{}_\beta &=& 0.\label{second2}
\end{eqnarray}
For the above gravitational wave ansatz, we have verified that all the 
irreducible parts of the curvature are trivial except for the pure tensor
pieces (\ref{Z1}) and (\ref{W1}). Accordingly, the direct computation of 
the gravitational hypermomentum 3-form then yields
\begin{equation}
H^\alpha{}_\beta = {}^\ast\!\left(w_1\,{}^{(1)}W^\alpha{}_\beta 
+ z_1\,{}^{(1)}Z^\alpha{}_\beta\right).\label{Hab}
\end{equation}

Let us now demonstrate that the gravitational wave ansatz above provides
an exact solution for the the gravitational field equations (\ref{first2})
and (\ref{second2}). The following two facts are crucial for this. The first 
one is the property of the curvature 2-form (\ref{curv}) 
\begin{equation}\label{kR}
k_\alpha R^\alpha{}_\beta = 0,\qquad k^\beta R^\alpha{}_\beta = 0,
\end{equation}
which is obviously satisfied due to the null nature of the wave vector
$k$ and its orthogonality to the 1-form $\gamma_\alpha$. 
The second fact concerns the well-known double duality property for the 
first irreducible part of the curvature:
\begin{equation}
{}^{\ast (1)}W_{\alpha\beta}= {\frac 1 2}\eta_{\alpha\beta\mu\nu}
\,{}^{(1)}W^{\mu\nu},\label{ddW1}
\end{equation}

We begin with the first equation (\ref{first2}). The property (\ref{kR})
obviously yields for the gravitational wave configuration 
$V_{\rm MAG}= -\,{\frac{1}{2}}\,R^{\alpha\beta}\wedge H_{\alpha\beta} = 0$
as well as the contraction $(e_{\alpha}\rfloor R_{\beta}{}^{\gamma})\wedge 
H^{\beta}{}_{\gamma} = 0$. Hence, our ansatz solves the first equation.

Finally, we turn to the second equation (\ref{second2}). Substituting 
(\ref{Hab}), we notice that the second term vanishes,
\begin{equation}
z_1\,D\,{}^\ast\!\left({}^{(1)}Z^\alpha{}_\beta\right) = 
z_1\,k^\alpha k_\beta\,d\,{}^\ast du = 0,
\end{equation}
due to the property (\ref{Dk}) and the radiation conditions (\ref{ddu}).
As a result, the second equation (\ref{second2}) reduces to
$w_1\,D\,{}^\ast\!\left({}^{(1)}W^\alpha{}_\beta\right) = 0$. Since
${}^{(1)}W^\alpha{}_\beta =  g^{\alpha\gamma}\,{}^{(1)}W_{\gamma\beta}$,
we have $D\,{}^\ast\!\left({}^{(1)}W^\alpha{}_\beta\right) = g^{\alpha\gamma}
\,D\,{}^\ast\!\left({}^{(1)}W_{\gamma\beta}\right) + Q^{\alpha\gamma}\wedge
{}^\ast\!\left({}^{(1)}W_{\gamma\beta}\right)$. Using the explicit form the 
nonmetricity (\ref{nonm}), we prove that the last term vanishes. Thus we
find, in the end,
\begin{equation}
w_1\,D\,{}^\ast\!\left({}^{(1)}W_{\alpha\beta}\right) = 0.
\end{equation}
Now we can use the double duality identity (\ref{ddW1}) and obtain
\begin{equation}\label{dW}
{\frac {w_1} 2}\,\eta_{\alpha\beta\mu\nu}\,D\,{}^{(1)}W^{\mu\nu} = 0,
\end{equation}
where we used the fact that the covariant derivative of the Levi-Civita 
tensor $D\eta_{\alpha\beta\mu\nu} = -\,2Q\,\eta_{\alpha\beta\mu\nu} = 0$
vanishes due to the absence of the Weyl covector, $Q = {\frac 14}
Q^\alpha{}_\alpha = 0$.

In order to demonstrate that our ansatz solves (\ref{dW}), we can use the 
MAG Bianchi identity which reads $DR_\alpha{}^\beta = 0$. We find $D\,
(g_{\alpha\gamma}\,R^{\gamma\beta}) = g_{\alpha\gamma}D R^{\gamma\beta} 
- Q_{\alpha\gamma}\wedge R^{\gamma\beta}$. Because of (\ref{nonm}) and 
(\ref{curv}), the last term vanishes for the gravitational wave 
configuration. Thus, in view of the Bianchi identity, the gravitational
wave curvature satisfies $D R^{\alpha\beta} = 0$. It now remains to use
the explicit formulas (\ref{curv}), (\ref{WZ}), (\ref{Z1}) and (\ref{W1}) 
to verify that 
\begin{equation}
D{}^{(1)}W^{\alpha\beta} = 0,
\end{equation}
since the covariant derivative of the symmetric curvature vanishes 
identically $D{}^{(1)}Z^{\alpha\beta} = k^\alpha k^\beta\,ddu \equiv 0$. 

Consequently, (\ref{dW}) is satisfied by the gravitational wave ansatz, and
this completes the proof that such a configuration is, indeed, an exact 
solution of the MAG field equations.

\section{Discussion and conclusion}\label{conclusion}

The extension of the Riemannian geometry of Einstein's general relativity
to the post-Riemannian structures of the metric-affine gravity can be 
motivated by a number of reasons. Among them, we mention the problem of
quantization (see the discussion of the renormalizable MAG models in  
\cite{lee1,lee2}), the theory of defects in the continuous media with 
microstructure (for a overview, see \cite{frank} and \cite{PR}), the 
physics of hadrons in terms of extended structures (see \cite{nee1,nee2,nee3}
and more details and references in \cite{PR}), the study of the early 
universe (in particular, relating the post-Riemannian structures to the
dark matter problem, see \cite{dirk3,dirk4,dirk5}). Finally, one can show
that the MAG models may arise as the effective theories in the context of
the dilaton-axion-metric low-energy limit of the string theory (see, e.g.,
\cite{dil1,dil2,dil3,dil4}). The study of the exact solutions of the MAG 
field equations is important for understanding and development of the 
physical aspects mentioned above. 

In this paper, we have derived a new plane wave solution of the general 
Yang-Mills type (curvature quadratic) metric-affine theory of gravity. This
extends the previous study of the waves in the Yang-Mills type models of 
the Poincar\'e gauge gravity \cite{adam,chen,sippel,vadim,singh,babu}. As 
compared to the other exact wave solutions available in the literature 
\cite{dirk1,dirk2,king,vas1,vas2,vas3}, the new configuration has
the following characteristic properties: (i) the spacetime metric 
is not a flat Minkowski one but the metric of the Riemannian gravitational
plane wave determined by the single harmonic function $H(\sigma, z^a)$,
(ii) there are not only torsion waves present but the nonmetricity has 
a nontrivial wave behavior as well, (iii) the post-Riemannian sector 
of the torsion and nonmetricity does not belong to the triplet ansatz. 
It is worthwhile to note that the triplet ansatz might be considered as
a useful tool which helps to avoid a possible problem of the well-posedeness
of the field equations by reducing them to the effective Einstein-Maxwell
system of equations. However, this ansatz is applicable only to a quite 
narrow class of the MAG models, namely to those Lagrangians (\ref{QMA}) 
where the only nontrivial coupling constant $z_4\neq 0$ is allowed. Our 
results apply to the general case with all the 16 nontrivial coupling 
constants $w_1, \ldots w_7$, $z_1, \ldots z_9$. The well-posedness of the 
general MAG model was never studied in the literature, and this question
clearly represents an open potentially interesting and important problem 
within the metric-affine approach to gravity. 

The results obtained have a number of interesting mathematical and physical 
applications. To begin with, the curvature quadratic Lagrangian is 
potentially important for a quantized theory of gravity. Furthermore, the
long-distance character of the wave solutions makes them a convenient tool
for the tests of the additional properties of matter besides the mass
(energy-momentum), namely, the hypermomentum which includes the spin and
the dilaton/shear charges. This is of particular interest for the study 
of the elastic media with defects, and for the physics of hadrons (see the
references quoted above).   

It is worthwhile to stress that the new solutions are obtained as the direct 
generalization of the general-relativistic wave solutions. When the 1-form
$u$ vanishes, the post-Riemannian geometric quantities disappear. On
the other hand, the Riemannian (metric) sector of the solution has the same
form as in general relativity, and thus all the earlier mathematical and 
physical analyses \cite{peres,pen1,pen2,griff,vdz,exact} are directly 
applicable to our case. In physical terms this means that the usual general 
relativistic detectors (with mass as the only gravitational charge) will not
distinguish between the gravitational waves of the Einstein theory and 
the new MAG waves. This is in complete agreement with the correspondence 
principle which underlies the dynamical structure of the metric-affine gravity:
MAG is not supposed to replace the general relativity theory in the well
established macroscopic domain, but rather to extend the latter in the 
microscopic domain by taking into account the additional physical properties
of matter (such as the spin, dilation and hypermomentum currents). The above
conclusion is based on the fact that the equations of motion in MAG for the 
test particles without the spin, dilaton and proper hypermomentum exactly 
coincide with the equations of motion of the test massive matter in general 
relativity \cite{nee}. Furthermore, it is worthwhile to recall that within 
the Poincar\'e gauge gravity the equations of motion of matter are also known 
to coincide with the general-relativistic equations of motion for the bodies 
with the trivial average spin value \cite{yass}. The corresponding 
generalization is expected to be valid in MAG for the macroscopic bodies 
with vanishing average spin, dilaton and proper hypermomentum, although 
the detailed relevant analysis is still missing in the literature. 

The new solution gives a natural generalization of the definitions of 
a gravitational plane-fronted wave. In accordance with \cite{orr}, 
a gravitational wave is defined by the existence of a null, geodetic, 
shear-, twist, and expansion-free vector field $k$ and the Weyl 2-form 
subject to the algebraic conditions
\begin{equation}
{}^{(1)}W^{\alpha\beta}\,k_\beta = 0,\qquad 
{}^{(1)}W^{[\alpha\beta}\,k^{\gamma]} =0.\label{wavedef1}
\end{equation}
According to \cite{sippel}, a gravitational wave is defined by a 
covariantly constant null field and a quadratic algebraic condition on
the components of the curvature tensor
\begin{equation}
R_{\mu\nu\alpha}{}^\beta\,R_{\rho\sigma\beta}{}^\alpha = 0.\label{wavedef2}
\end{equation}
As we can immediately verify, the non-Riemannian curvature (as well as its
Riemannian constituent) satisfies both (\ref{wavedef1}) and (\ref{wavedef2}).

In \cite{king,vas1,vas2,vas3} the notion of the pseudo-instanton solutions
of the MAG field equations was introduced. The latter are described by
a metric-compatible linear connection for which only one of the 
eleven irreducible parts of the curvature is nontrivial. Our gravitational
wave solution provides a minimal generalization of the pseudo-instanton, 
in the sense that the nonmetricity does not vanish and that the curvature
has {\it two} purely tensor irreducible parts (\ref{Z1}) and (\ref{W1}).

The applications mentioned above refer to the fundamental MAG theory. However,
we recall that MAG also arises as an effective theory within the framework of
the dilaton-axion-metric low energy limit of the string models. Accordingly, 
one can use our solution as a technical tool to construct the exact wave 
configurations in the string motivated models where the plane waves play 
essential role \cite{gimon,ark1,ark2,str1,str2,str3,str4}. This construction 
will be described in detail elsewhere.  

\bigskip
{\bf Acknowledgment} This work was supported by the Deutsche 
Forschungsgemeinschaft (Bonn) with the grant HE~528/20-1. I thank 
Friedrich Hehl for the reading of the manuscript and for the discussion 
of the results obtained.


\begin{thebibliography}{99}
\bibitem{remark}
Non-wave solutions were studied \cite{eff} for a class of more general 
quadratic MAG models with the Lagrangian which includes the Hilbert-Einstein 
term plus all possible quadratic invariants constructed from the torsion and 
nonmetricity, for an overview see \cite{tri}.
\end{thebibliography}
\end{document}